\documentclass{llncs}

\usepackage{amssymb}
\usepackage{amsmath}
\usepackage{color}
\usepackage{graphics,latexsym,amsfonts}
\usepackage{graphicx}
\usepackage{eurosym}
\usepackage{multicol}
\usepackage{pstricks, pst-node}

\newcommand{\IU}{\mathbb{U}}
\newcommand{\IR}{\mathbb{R}}

\advance\textwidth by 2.5truecm
\advance\oddsidemargin by -1truecm
\advance\textheight by 5.5truecm \advance\topmargin by -2.5truecm

\begin{document}
\title{Asymmetric R\&D Alliances and Coopetitive Games}

\author{Daniela Baglieri*, David Carf\`{i}**, Giovanni Battista Dagnino***}

\institute{*University of Messina\\
Via dei Verdi 75, 98122, Messina, Italy.\\
**Department of Mathematics\\
University of California at Riverside, USA\\
%900 Big Springs Road, Surge 231 Riverside \\
%CA 92521-0135, USA.\\
***University of Catania \\
Corso Italia 55, 95129, Catania, Italy.\\
 \email{**dbaglieri@unime.it}\\
 \email{**davidcarfi71@yahoo.it}\\
 \email{***dagnino@unict.it}
 }

%\date{29 may 07}
\date{\today}

\maketitle

%%%%%%%%%%%%%%%%%%%%%%%%%%%%%%%%%%%%%%%%%%%%%%%%%%%%%%%%%%%%%%%%
%
%                   ABSTRACT
%                                                                                                                                            
%%%%%%%%%%%%%%%%%%%%%%%%%%%%%%%%%%%%%%%%%%%%%%%%%%%%%%%%%%%%%%%

\begin{abstract}

\noindent 
In this paper we show how the study of asymmetric R\&D alliances, that are those between young and small firms and large and MNEs firms for knowledge exploration and/or exploitation, requires the adoption of a coopetitive framework which consider both collaboration and competition. We draw upon the literature on asymmetric R\&D collaboration and coopetition to propose a mathematical model for the coopetitive games which is particularly suitable for exploring asymmetric R\&D alliances.

\medskip


\end{abstract}
%%%%%%%%%%%%%%%%%%%%%%%%%%%%%%%%%%%%%%%%%%%%%%%%%%%%%%%%%%%%%%%%%%%%%%%%%%%%%%%%%%%%%%%%%%%%%%%%%%%%%%%%%%%%%%%%%%%%%%%%%%

\section{Introduction}

Scholarly attention to co-opetition has increased with the practical significance of collaboration among competitors (Brandenburger and Nalebuff, 1996; Sakakibara, M. 1997;  Padula and Dagnino, 2007) and competition among \emph{friends} (for instance, university-industry relationships; see: Carayannis and Alexander, 1999; Baglieri 2009). 

\textbf{R\&D alliances.} Despite the increased importance of co-opetition, limited research has examined factors that may drive co-opetition, particularly in high technology industries where R\&D alliances seem to be growing rapidly. A notable trend is the rapid growth of R\&D alliances between large, well-established firms and small, growing firms. We term these alliances asymmetric R\&D alliances. These inter-organizational arrangements rise a number of open questions related to the disparately partners bargaining power which affect, among others, alliances outcomes.

(1) Are asymmetric R\&D alliances a win-win or win-lose partnerships? 

(2) What are the main firms' strategies partners may deploy to \emph{enlarge the pie} and create more value?

The answers to these questions are important for both larger and smaller firms to better select their partners, the scope and type of alliance, and the resources to be allocated for new product development. 

\textbf{Coopetitive games.} This paper aims at developing a mathematical model for the coopetitive games which is particularly suitable for exploring asymmetric R\&D alliances. Despite the classic form games involving two players - that can choose the respective strategies only cooperatively or not-cooperatively, in an exclusive way - we propose that players have a common strategy set $C$, containing other strategies (possibly of different type with respect to the previous one) that must be chosen cooperatively. Consequently, for any coopetitive game, we necessarily build up a family of classic normal-form games which determines univocally the given coopetitive game. Thus, the study of a coopetitive game is reconducted to the study of a family of normal-form games in its completeness. In this paper, we suggest how this latter study can be conduct and what could be the concepts of solution of a coopetitive game corresponding to the main firms' strategies, potentially deployed in asymmetric R\&D settings.

\textbf{Asymmetric R\&D alliances: a coopetitive perspective.} Several researchers have clearly indicated the importance of co-opetition for technological innovation.

(1) Jorde and Teece (1990) suggested that the changing dynamics of technologies and markets have led to the emergence of the simultaneous innovation model. For firms to pursue the simultaneous innovation model and succeed in innovation, they should look for collaboration opportunities that allow them to bring multiple technologies and diverse and complementary assets together. 

(2) Von Hippel (1987) argued that collaboration for knowledge sharing among competitors occurs when technological progress may be faster with collective efforts rather than through individual efforts and when combined knowledge offers better advantages than solo knowledge. More recent research clearly shows the importance of co-opetition in technological innovation. 

(3) Quintana-Garcia and Benavides-Velasco (2004) empirically show that collaboration with direct competitors is important not only to acquire new technological knowledge and skills from the partner, but also to create and access other capabilities based on intensive exploitation of the existing ones.

(4) Similarly, Carayannis and Alexander (1999) argue that co-opetition is particularly important in knowledge intensive, highly complex, and dynamic environments.

% scarti %
	
\textbf{Coopetition among firms.} Following the seminal work of Nalebuff and Brandenburger (1996), some scholars have studied how firms cooperate in the upstream activities and compete in the downstream activities (Walley, 2007), in order to pursuit several goals:

(1) to integrate complementary resources within the value net (Nalebuff and Brandenburger, 1996; Bagshaw and Bagshaw, 2001; Wilkinson and Young, 2002; Laine, 2002);

(2) to increase the heterogeneity of the resources needed to successfully compete in convergent businesses (Ancarani and Costabile, 2006);

(3) to enhance learning opportunities (Hamel et alii, 1989);

(4) to boost firmÕs R\&D capabilities (Valentini et alii, 2004);

(5) to speed up innovation (Hagel III and Brown, 2005).

\smallskip
	
\begin{small}
 
\textbf{Note.} This broad range of goals explains why coopetition is common in several industries:

(1) Sakakibara (1993) describes R\&D cooperation among competing Japanese semiconductor firms.

(2) Hagedoorn, Cyrayannis, and Alexander (2001) document an emerging collaboration between IBM and Apple that resulted in an increasing number of alliances between the two for joint technological development.

(3) Albert (1999) points to the coopetitive relationship between Dell Computers and IBM.

(4) Coopetition is common in mature industries too.

(5) Recent works examine coopetition in soft drink and beverage industry (Bonel and Rocco, 2007); in carbonated soft drink industry (Fosfuri and Giarratana, 2006); and in tuna industry (LeRoy, 2006). 

Therefore, the idea to Òworking with the enemyÓ is not new, although it has been a \emph{an under researched theme} (Dagnino and Padula, 2002).

\end{small}
	
\textbf{What is new here?} What it is new in this paper is the attempt to apply coopetition in a asymmetric R\&D alliances setting, theme which has been more investigated in industrial organizational literature that largely explores how increases in cooperative activity in the markets for technology (i.e. licensing) affect levels of competitive activity in Òproduct marketsÓ. In this paper, we adopt another option which addresses the question how firms manage simultaneously patterns of cooperation and competition in their R\&D relationships (firm level) and, thus, how firms Òenlarge the pieÓ (cooperation) and Òshare the pieÓ (competition). According to a process view, we propose a mathematical model to determine possible suitable behaviors (actions) of partners during their strategic interactions, from both non-cooperative and cooperative point of view.

\textbf{Organization of the paper.} In the first section we introduce the model of coopetitive game of D. Carf\`{i}. In the second we apply the complete analysis of a differentiable game (see \cite {ca1,ca3,ca5}) to an asymmetric interaction of two firms. In the last section we study a possible coopetitive extension of that asymmetric interaction to obtain a win-win situation in a R\&D coopetitive perspective. 

\section{Coopetitive games}

In this paper we show and apply the mathematical model of a \emph{
coopetitive game} introduced by David Carf\`{i} in \cite{ca2} and  \cite{ca4}. The idea of
coopetitive game is already used, in a mostly intuitive and non-formalized
way, in Strategic Management Studies (see for example Brandenburgher and Nalebuff).

\smallskip

\textbf{The idea.} A coopetitive game is a game in which two or more players (participants)
can interact \emph{cooperatively and non-cooperatively at the same time}.
Even Brandenburger and Nalebuff, creators of coopetition, did not
define, precisely, a \emph{quantitative way to implement coopetition} in the Game
Theory context.

\smallskip

The problem to implement the notion of coopetition in Game Theory is summarized in the following question:

\emph{how do, in normal form games, cooperative and non-cooperative interactions can live together simultaneously, in a Brandenburger-Nalebuff sense?}

\smallskip

To explain the above question, consider a classic two-player normal-form gain game $G = (f,>)$ - such
a game is a pair in which $f$ is a vector valued function defined on a
Cartesian product $E\times F$ with values in the Euclidean plane $\Bbb{R}^{2}$ and $>$ is the natural strict sup-order of the Euclidean plane itself. Let $E$ and $F$ be the strategy sets of the two players in the game $G$. The two players
can choose the respective strategies $x\in E$ and $y\in F$ cooperatively
(exchanging information) or not-cooperatively (not exchanging informations),
but these two behavioral ways are mutually exclusive, at least in normal-form games: the two ways
cannot be adopted simultaneously  in the model of normal-form
game (without using convex probability mixtures, but this is not the
way suggested by Brandenburger and Nalebuff in their approach). There is no room, in the classic normal game model, for a simultaneous (non-probabilistic) employment of the two behavioral extremes \emph{cooperation} and 
\emph{non-cooperation}.

\textbf{Towards a possible solution.} David Carf\`{i} (\cite{ca2} and  \cite{ca4}) has
proposed a manner to pass this \emph{impasse}, according to the idea of
coopetition in the sense of Brandenburger and Nalebuff:

in a Carf\`{i}'s coopetitive game model, the players of the game have their
respective strategy-sets (in which they can choose cooperatively or not
cooperatively) and a common strategy set $C$ containing other strategies
(possibly of different type with respect to those in the respective classic strategy sets) that \emph{must be chosen cooperatively}. This strategy set $C$ can also be structured
as a Cartesian product (similarly to the profile strategy space of normal form games), but in any case the strategies belonging to this new
set $C$ must be chosen cooperatively.

\subsection{The model for $n$ players}

We give in the following the definition of coopetitive game proposed by
Carf\`{i} (in \cite{ca2} and  \cite{ca4}).

\smallskip

\textbf{Definition (of }$n$\textbf{-player coopetitive game).}\emph{ Let $E=(E_{i})_{i=1}^{n} $ be a finite $n$-family of non-empty sets and let $C$ be another non-empty set. We define $n$-\textbf{player coopetitive
gain game over the strategy support }$(E,C)$ any pair $G=(f,>)$,
where $f$ is a vector function from the Cartesian product $^{\times }E\times C$ (here $^{\times }E$ denotes the classic
strategy-profile space of $n$-player normal form games, i.e. the
Cartesian product of the family $E$) into the $n$-dimensional
Euclidean space $\Bbb{R}^{n}$ and $>$ is the natural
sup-order of this last Euclidean space. The element of the set $C$ will be called \textbf{cooperative strategies of the game}.}

\smallskip

A particular aspect of our coopetitive game model is that any coopetitive game $G$ determines univocally a family of classic normal-form games and vice versa; so that any coopetitive game could be defined as a family of normal-form games. In what follows we precise this very important aspect of the model.

\smallskip

\textbf{Definition (the family of normal-form games associated with a
coopetitive game).}\emph{\ Let }$G=(f,>)$\emph{\ be a coopetitive game over
a strategic support }$(E,C)$\emph{. And let} $g = (g_{z})_{z\in C}$ \emph{be the family of classic normal-form games whose member }$g_{z}$\emph{\ is, for any cooperative strategy }$z$\emph{\ in }$C$\emph{, the normal-form game} $G_{z}:=(f(.,z),>),$ \emph{where the payoff function }$f(.,z)$\emph{\ is the section} $ f(.,z):\;^{\times }E\rightarrow \Bbb{R}^{n}$ \emph{of the function $f$, defined (as usual) by} $f(.,z)(x)=f(x,z),$ \emph{for every point }$x$\emph{\ in the strategy profile space }$^{\times}E $\emph{. We call the family }$g$\emph{\ (so defined) \textbf{family of
normal-form games associated with (or determined by) the game} }$G$\emph{ and we call \textbf{normal section} of the game $G$ any member of the family $g$.}

\smallskip

We can prove this (obvious) theorem.

\smallskip

\textbf{Theorem.}\emph{\ The family $g$ of normal-form games
associated with a coopetitive game $G$ uniquely determines the game.
In more rigorous and complete terms, the correspondence $G\mapsto g$
is a bijection of the space of all coopetitive games - over the strategy
support $(E,C)$ - onto the space of all families of normal form
games - over the strategy support $E$ - indexed by the set $C$.}

%\emph{Proof.} This depends on the fact that we have the following
%natural bijection between function spaces: $\mathcal{F}(^{\times }E\times C,\Bbb{R}^{n})\rightarrow \mathcal{F}(C,\mathcal{F}(^{\times }E,\Bbb{R}^{n})):f\mapsto (f(.,z))_{z\in C}$; which is a classic result of theory of sets. $\blacksquare $

\smallskip

Thus, the exam of a coopetitive game should be equivalent to the exam of a
whole family of normal-form games (in some sense we shall specify). In this paper we suggest how this latter examination can be conducted and what are the solutions corresponding to the main concepts of solution which
are known in the literature for the classic normal-form games, in the case of
two-player coopetitive games.

\subsection{Two players coopetitive games}

In this section we specify the definition and related concepts of two-player
coopetitive games; sometimes (for completeness) we shall repeat some
definitions of the preceding section.

\smallskip

\textbf{Definition.}\emph{\ Let }$E$\emph{, }$F$\emph{
\ and }$C$\emph{\ be three nonempty sets. We define \textbf{two player
coopetitive gain game carried by the strategic triple} }$(E,F,C)$\emph{\ any
pair of the form} $G=(f,>),$ \emph{where }$f$\emph{\ is a function from the Cartesian product }$E\times
F\times C$\emph{\ into the real Euclidean plane }$\Bbb{R}^{2}$\emph{\ and
the binary relation }$>$\emph{\ is the usual sup-order of the Cartesian
plane (defined component-wise, for every couple of points }$p$\emph{\ and }$q
$\emph{, by }$p>q$\emph{\ iff }$p_{i}>q_{i}$\emph{, for each index }$i$\emph{%
).}

\smallskip

\begin{small}

\textbf{Remark (coopetitive games and normal form games).} The difference
among a two-player normal-form (gain) game and a two player coopetitive (gain) game
is the fundamental presence of the third strategy Cartesian-factor $C$. The
presence of this third set $C$ determines a total change of perspective with
respect to the usual exam of two-player normal form games, since we now have to consider
a normal form game $G(z)$, for every element $z$ of the set $C$; we have,
then, to study an entire ordered family of normal form games in its own
totality, and we have to define a new manner to study these kind of game
families.

\end{small}
 
\subsection{\textbf{Terminology and notation}}

\textbf{Definitions.}\emph{\ Let }$G=(f,>)$\emph{\ be a two player
coopetitive gain game carried by the strategic triple }$(E,F,C)$\emph{. We
will use the following terminologies:}

1.  \emph{the function }$f$\emph{\ is called the \textbf{payoff function
of the game} }$G$\emph{;}

2.  \emph{the first component }$f_{1}$\emph{\ of the payoff function }$f$%
\emph{\ is called \textbf{payoff function of the first player} and
analogously the second component }$f_{2}$\emph{\ is called \textbf{payoff
function of the second player};}

3.  \emph{the set }$E$\emph{\ is said \textbf{strategy set of the first
player} and the set }$F$\emph{\ the \textbf{strategy set of the second player%
};}

4.  \emph{the set }$C$\emph{\ is said the \textbf{cooperative
strategy set of the two players};}

5.  \emph{the Cartesian product }$E\times F\times C$\emph{\ is called the strategy space of the game} $G$\emph{.}

%\textbf{Memento.} The first component $f_{1}$ of the payoff function $f$ of
%a coopetitive game $G$ is the function of the strategy space $E\times
%F\times C$ of the game $G$ into the real line $\Bbb{R}$ defined by the first
%projection $f_{1}(x,y,z):=\mathrm{pr}_{1}(f(x,y,z))$,
%for every strategic triple $(x,y,z)$ in $E\times F\times C$; in a similar fashion we
%proceed for the second component $f_{2}$ of the function $f$.

\smallskip

\textbf{Interpretation.} We have:

1.  two players, or better an ordered pair $(1,2)$ of players;

2.  anyone of the two players has a strategy set in which to choose
freely his own strategy;

3.  the two players can/should \emph{cooperatively} choose strategies $z$
in a third common strategy set $C$;

4.  the two players will choose (after the exam of the entire game $G$)
their cooperative strategy $z$ in order to maximize (in some sense we shall
define) the vector gain function $f$.

\subsection{\textbf{Normal form games of a coopetitive game}}

Let $G$ be a coopetitive game in the sense of above definitions. For any
cooperative strategy $z$ selected in the cooperative strategy space $C$,
there is a corresponding normal form gain game $G_{z}=(p(z),>)$, upon the strategy pair $(E,F)$, where the payoff function $p(z)$ is the section $f(.,z):E\times F\to \Bbb{R}^{2}$, of the payoff function $f$ of the coopetitive game - the section is defined,
as usual, on the competitive strategy space $E\times F$, by $f(.,z)(x,y)=f(x,y,z)$,
for every bi-strategy $(x,y)$ in the bi-strategy space $E\times F$.

\smallskip

Let us formalize the concept of game-family associated with a coopetitive
game.

\smallskip

\textbf{Definition (the family associated with a coopetitive game).}\emph{\
Let }$G=(f,>)$\emph{\ be a two player coopetitive gain game carried by the
strategic triple }$(E,F,C)$\emph{. We naturally can associate with the game }%
$G$\emph{\ a family }$g=(g_{z})_{z\in C}$ \emph{of normal-form games defined
by $g_{z} := G_{z}=(f(.,z),>)$, for every }$z$ \emph{in }$C$\emph{, which we shall call \textbf{the family
of normal-form games associated with the coopetitive game} }$G$\emph{.}

\smallskip

\textbf{Remark.} It is clear that with any above family of normal form games $g=(g_{z})_{z\in C}$, with $g_{z}=(f(.,z),>)$, we can associate:

1.  a family of payoff spaces $(\mathrm{im}f(.,z))_{z\in C}$, with members in the payoff universe $\Bbb{R}^{2}$;

2.  a family of Pareto maximal boundary $(\partial ^{*}G_{z})_{z\in C}$, with members contained in the payoff universe $\Bbb{R}^{2}$;

3.  a family of suprema $(\mathrm{sup}G_{z})_{z\in C}$, with members belonging to the payoff universe $\Bbb{R}^{2}$;

4.  a family of Nash zones $(\mathcal{N}(G_{z}))_{z\in C}$;
with members contained in the strategy space $E\times F$;

5.  a family of conservative bi-values $v^{\#}=(v_{z}^{\#})_{z\in C}$;
in the payoff universe $\Bbb{R}^{2}$.

And so on, for every meaningful known feature of a normal form game. Moreover, we can interpret any of the above families as \emph{set-valued paths} in
the strategy space $E\times F$ or in the payoff universe $\Bbb{R}^{2}$. It is just the study of these induced families which becomes of great
interest in the examination of a coopetitive game $G$ and which will enable
us to define (or suggest) the various possible solutions of a coopetitive game.

\section{Solutions of a coopetitive game}

The two players of a coopetitive game $G$ should choose the cooperative
strategy $z$ in $C$ in order that:

1.  the reasonable Nash equilibria of the game $G_{z}$ are $f$-preferable
than the reasonable Nash equilibria in each other game $G_{z^{\prime }}$;

2.  the supremum of $G_{z}$ is greater (in the sense of the usual order
of the Cartesian plane) than the supremum of any other game $G_{z^{\prime }}$;

3.  the Pareto maximal boundary of $G_{z}$ is higher than that of any
other game $G_{z^{\prime }}$;

4.  the Nash bargaining solutions in $G_{z}$ are $f$-preferable than
those in $G_{z^{\prime }}$;

5.  \emph{in general, fixed a common kind of solution for any game }$%
G_{z} $\emph{, say }$S(z)$\emph{\ the set of these kind of solutions for the
game }$G_{z}$\emph{, we can consider the problem to find all the optimal
solutions (in the sense of Pareto) of the set valued path }$S$\emph{,
defined on the cooperative strategy set }$C$\emph{. Then, we should face the
problem of \textbf{selection of reasonable Pareto strategies} in the
set-valued path }$S $\emph{\ via proper selection methods (Nash-bargaining,
Kalai-Smorodinsky and so on).}

Moreover, we shall consider the maximal Pareto boundary of the payoff space $\mathrm{im}(f)$ as an appropriate zone for the bargaining solutions. The payoff function of a two person coopetitive game is (as in the case of
normal-form game) a vector valued function with values belonging to the
Cartesian plane $\Bbb{R}^{2}$. We note that in general the above criteria
are multi-criteria and so they will generate multi-criteria optimization
problems. In this section we shall define rigorously some kind of solution, for two
player coopetitive games, based on a bargaining method, namely a
Kalai-Smorodinsky bargaining type. Hence, first of all, we have to precise
what kind of bargaining method we are going to use.

\subsubsection{Bargaining problems.}

In this paper we shall use the following definition of bargaining problem.

\smallskip

\textbf{Definition (of bargaining problem).}\emph{\ Let }$S$\emph{\ be a
subset of the Cartesian plane }$\Bbb{R}^{2}$\emph{\ and let }$a$\emph{\ and }%
$b$\emph{\ be two points of the plane with the following properties:}

1.  \emph{they belong to the small interval containing }$S$\emph{, if this interval is defined (indeed, it is well defined if and only if $S$ is bounded and it is precisely the interval $[\inf S,\sup S]_{^{\leq }}$);}

2.  \emph{they are such that }$a<b$\emph{;}

3.  \emph{the intersection} $[a,b]_{^{\leq }}\cap \partial ^{*}S$, \emph{among the interval }$[a,b]_{^{\leq }}$\emph{\ with end points }$a$ \emph{\ and }$b$\emph{\ (it is the set of points greater than }$a$\emph{\
and less than }$b$\emph{, \textbf{it is not} the segment }$[a,b]$\emph{) and
the maximal boundary of }$S$ \emph{is non-empty.}

\emph{In this conditions, we call\textbf{\ bargaining problem on} }$S$\emph{%
\ \textbf{corresponding to the pair of extreme points} }$(a,b)$\emph{, the
pair} $P=(S,(a,b))$. \emph{Every point in the intersection among the interval }$[a,b]_{^{\leq }}$%
\emph{\ and the Pareto maximal boundary of }$S$\emph{\ is called \textbf{%
possible solution of the problem} }$P$\emph{. Some time the first extreme
point of a bargaining problem is called \textbf{the initial point of the problem}
(or \textbf{disagreement point} or \textbf{threat point}) and the second
extreme point of a bargaining problem is called \textbf{utopia point} of the problem.}

\smallskip

In the above conditions, when $S$ is convex, the problem $P$ is said to be
convex and for this case we can find in the literature many existence
results for solutions of $P$ enjoying prescribed properties
(Kalai-Smorodinsky solutions, Nash bargaining solutions and so on ...).

\smallskip

\textbf{Remark.} Let $S$ be a subset of the Cartesian plane $\Bbb{R}^{2}$
and let $a$ and $b$ two points of the plane belonging to the smallest
interval containing $S$ and such that $a\leq b$. Assume the Pareto maximal
boundary of $S$ be non-empty. If $a$ and $b$ are a lower bound and an upper
bound of the maximal Pareto boundary, respectively, then the intersection $[a,b]_{^{\leq }}\cap \partial ^{*}S$
is obviously not empty. In particular, if $a$ and $b$ are the extrema of $S$
(or the extrema of the Pareto boundary $S^{*}=\partial ^{*}S$) we can
consider the following bargaining problem $P=(S,(a,b))$ (or $P=(S^{*},(a,b))$) and we call this particular problem a \emph{standard bargaining problem on }$S$ (or \emph{standard bargaining problem on the Pareto maximal boundary} $S^{*}$).

\subsubsection{Kalai solution for bargaining problems.}

Note the following property.

\smallskip

\textbf{Property.}\emph{\ If }$(S,(a,b))$\emph{\ is a bargaining problem
with }$a<b$\emph{, then there is at most one point in the intersection} $[a,b]\cap \partial ^{*}S$, \emph{where }$[a,b]$\emph{\ is the \textbf{segment joining the two points} }$%
a$\emph{\ and }$b$\emph{.}

%\emph{Proof.} Since if a point $p$ of the segment $[a,b]$ belongs to the
%Pareto boundary $\partial ^{*}S$, no other point of the segment itself can
%belong to Pareto boundary, since the segment is a totally ordered subset of
%the plane (remember that $a<b$). $\blacksquare $

\smallskip

\textbf{Definition (Kalai-Smorodinsky). }\emph{We call \textbf{%
Kalai-Smorodinsky solution} (or \textbf{best compromise solution}) \textbf{%
of the bargaining problem} }$(S,(a,b))$ \emph{the unique point of the
intersection} $[a,b]\cap \partial ^{*}S$, \emph{if this intersection is non empty.}

%\smallskip

%So, in the above conditions, the Kalai-Smorodinsky solution $k$ (if it
%exists) enjoys the following property: there is a real $r$ in $[0,1]$ such
%that $k=a+r(b-a)$ or $k-a=r(b-a)$, 
%hence $(k_2-a_2)/(k_1-a_1) = (b_2 - a_2)/(b_1 - a_1)$, if the above ratios are defined; these last equality is the \emph{characteristic property of Kalai-Smorodinsky solutions}.

\smallskip

We end the subsection with the following definition.

\smallskip

\textbf{Definition (of Pareto boundary). }\emph{We call \textbf{Pareto
boundary} every subset }$M$\emph{\ of an ordered space which has only
pairwise incomparable elements.}

\subsubsection{Nash (proper) solution of a coopetitive game.}

Let $N:=\mathcal{N}(G)$ be the union of the Nash-zone family of a
coopetitive game $G$, that is the union of the family $(\mathcal{N}(G_{z}))_{z\in C}$ of all Nash-zones of the game family $g=(g_{z})_{z\in C}$
associated to the coopetitive game $G$. We call \emph{Nash path of the game} 
$G$ the multi-valued path $z\mapsto \mathcal{N}(G_{z})$ and Nash zone of $G$ the trajectory $N$ of the above multi-path. Let $N^{*}$ be the Pareto maximal boundary of the Nash zone $N$. We can consider the
bargaining problem $P_{\mathcal{N}}=(N^{*},\inf N^{*},\sup N^{*})$.

\smallskip

\textbf{Definition.}\emph{\ If the above bargaining problem }$P_{\mathcal{N}%
} $\emph{\ has a Kalai-Smorodinsky solution }$k$\emph{, we say that }$k$%
\emph{\ is\textbf{\ the properly coopetitive solution of the coopetitive game%
} }$G$\emph{.}

\smallskip

The term ``properly coopetitive'' is clear:

\begin{itemize}
\item  \emph{this solution }$k$ \emph{is determined by cooperation on the
common strategy set }$C$\emph{\ and to be selfish (competitive in the Nash
sense) on the bi-strategy space }$E\times F$\emph{.}
\end{itemize}

\subsubsection{Bargaining solutions of a coopetitive game.}

It is possible, for coopetitive games, to define other kind of solutions,
which are not properly coopetitive, but realistic and sometime affordable.
These kind of solutions are, we can say, \emph{super-cooperative}. Let us show some of these kind of solutions.

\smallskip

Consider a coopetitive game $G$ and

1.  its Pareto maximal boundary $M$ and the corresponding pair of extrema 
$(a_{M},b_{M})$;

2.  the Nash zone $\mathcal{N}(G)$ of the game in the payoff space and
its extrema $(a_{N},b_{N})$;

3.  the conservative set-value $G^{\#}$ (the set of all conservative
values of the family $g$ associated with the coopetitive game $G$) and its
extrema $(a^{\#},b^{\#})$.

\emph{We call:}

1.  \emph{\textbf{Pareto compromise solution of the game} }$G$\emph{\ the
best compromise solution (K-S solution) of the problem } $(M,(a_{M},b_{M}))$,
\emph{if this solution exists;}

2.  \emph{\textbf{Nash-Pareto compromise solution of the game} }$G$\emph{%
\ the best compromise solution of the problem } $(M,(b_{N},b_{M}))$ \emph{if this solution exists;}

3.  \emph{\textbf{conservative-Pareto compromise solution of the game} }$
G $\emph{\ the best compromise of the problem} $(M,(b^{\#},b_{M}))$ \emph{if this solution exists.}

\subsubsection{Transferable utility solutions.}

Other possible compromises we suggest are the following. Consider \emph{the transferable utility Pareto boundary} $M$ \emph{of the coopetitive game} $G$, that is the set of all points $p$ in the Euclidean
plane (universe of payoffs), between the extrema of $G$, such that their
sum $^+(p):=p_{1}+p_{2}$ is equal to the maximum value of the addition $+$ of the real line $\Bbb{R}$ over the
payoff space $f(E\times F\times C)$ of the game $G$.

\smallskip

\textbf{Definition (TU Pareto solution).} \emph{We call \textbf{transferable
utility compromise solution of the coopetitive game} }$G$\emph{\ the
solution of any bargaining problem }$(M,(a,b))$\emph{, where}

1.  $a$\emph{\ and }$b$ \emph{are points of the smallest interval
containing the payoff space of }$G$

2.  $b$\emph{\ is a point strongly greater than }$a$\emph{;}

3.  $M$\emph{\ is the transferable utility Pareto boundary of the game }$G $\emph{;}

4.  \emph{the points }$a$\emph{\ and }$b$\emph{\ belong to different
half-planes determined by }$M$\emph{.}

Note that the above forth axiom is equivalent to require that the segment
joining the points $a$ and $b$ intersect $M$.

\subsubsection{Win-win solutions.}

In the applications, if the game $G$ has a member $G_{0}$ of its family
which can be considered as an ``initial game'' - in the sense that the
pre-coopetitive situation is represented by this normal form game $G_{0}$ -
the aims of our study (following the standard ideas on coopetitive
interactions) are: 1) to ``enlarge the pie''; 2) to share the pie in order to obtain a win-win solution with respect to the initial situation. So that we will choose as a threat point $a$ in TU problem $(M,(a,b))$ the supremum of the initial game $G_{0}$.

\smallskip

\textbf{Definition (of win-win solution).}\emph{\ Let }$(G,z_{0})$\emph{\ be
a \textbf{coopetitive game with an initial point}, that is a coopetitive
game }$G$\emph{\ with a fixed common strategy }$z_{0}$\emph{\ (of its common
strategy set }$C$\emph{). We call the game }$G_{z_{0}}$\emph{\ as \textbf{%
the initial game of} }$(G,z_{0})$\emph{. We call \textbf{win-win solution of
the game} }$(G,z_{0})$\emph{\ any strategy profile }$s=(x,y,z)$\emph{\ such
that the payoff of }$G$\emph{\ at }$s$\emph{\ is strictly greater than the
supremum }$L$ \emph{of the \textbf{payoff core} of the initial game }$%
G(z_{0})$\emph{.}

\smallskip

\textbf{Remark.} The payoff core of a normal form gain game $G$ is the portion of the Pareto maximal boundary $G^{*}$ of the game which is greater than the conservative bi-value of $G$. 

\smallskip

\textbf{Remark.} From an applicative point of view, the above requirement
(to be strictly greater than $L$) is very strong. More realistically, we can
consider as win-win solutions those strategy profiles which are strictly
greater than any reasonable solution of the initial game $G_{z_{0}}$.

\smallskip

\textbf{Remark.} In particular, observe that, if the collective payoff
function $^{+}(f)=f_{1}+f_{2}$ has a maximum (on the strategy profile space $S$) strictly greater than the
collective payoff $L_{1}+L_{2}$ at the supremum $L$ of the payoff core of
the game $G_{z_{0}}$, the portion $M(>L)$ of TU Pareto boundary $M$ which is greater than $L$ is
non-void and it is a segment. So that we can choose as a threat point $a$ in
our problem $(M,(a,b))$ the supremum $L$ of the payoff core of the initial
game $G_{0}$ \emph{to obtain some compromise solution}.

\smallskip

\textbf{Standard win-win solution.} A natural choice for the utopia point $b$
is the supremum of the portion $M_{\geq a}$ of the transferable utility
Pareto boundary $M$ which is upon (greater than) this point $a$: $M_{\geq a}=\{m\in M:m\geq a\}$.

\section{An asymmetric interaction among firms}

\subsubsection{The economic situation.} We consider two economic agents, firms. The second one is already in the market the first one is not. The possible strategies of the first are Enter in the market (E) and Not Enter in market (N). The strategies of the second one are High prices (H)  and Low prices (L). The payoff (gains) of the two firms are represented in the following table:
\[
\begin{array}{ccc}

  	&	H	&  	L 	\\
  E	&	(4,2)	&   	(0,3)	\\
  N	&	(0,3)	&   	(0,4)		   

\end{array}
\]

\subsubsection{The finite game associated with the economic situation.} The above table defines a bimatrix $M$ and, consequently, a finite loss game $(M,<)$.

\begin{figure}[htbp]
\begin{center}
\includegraphics [width=6cm]{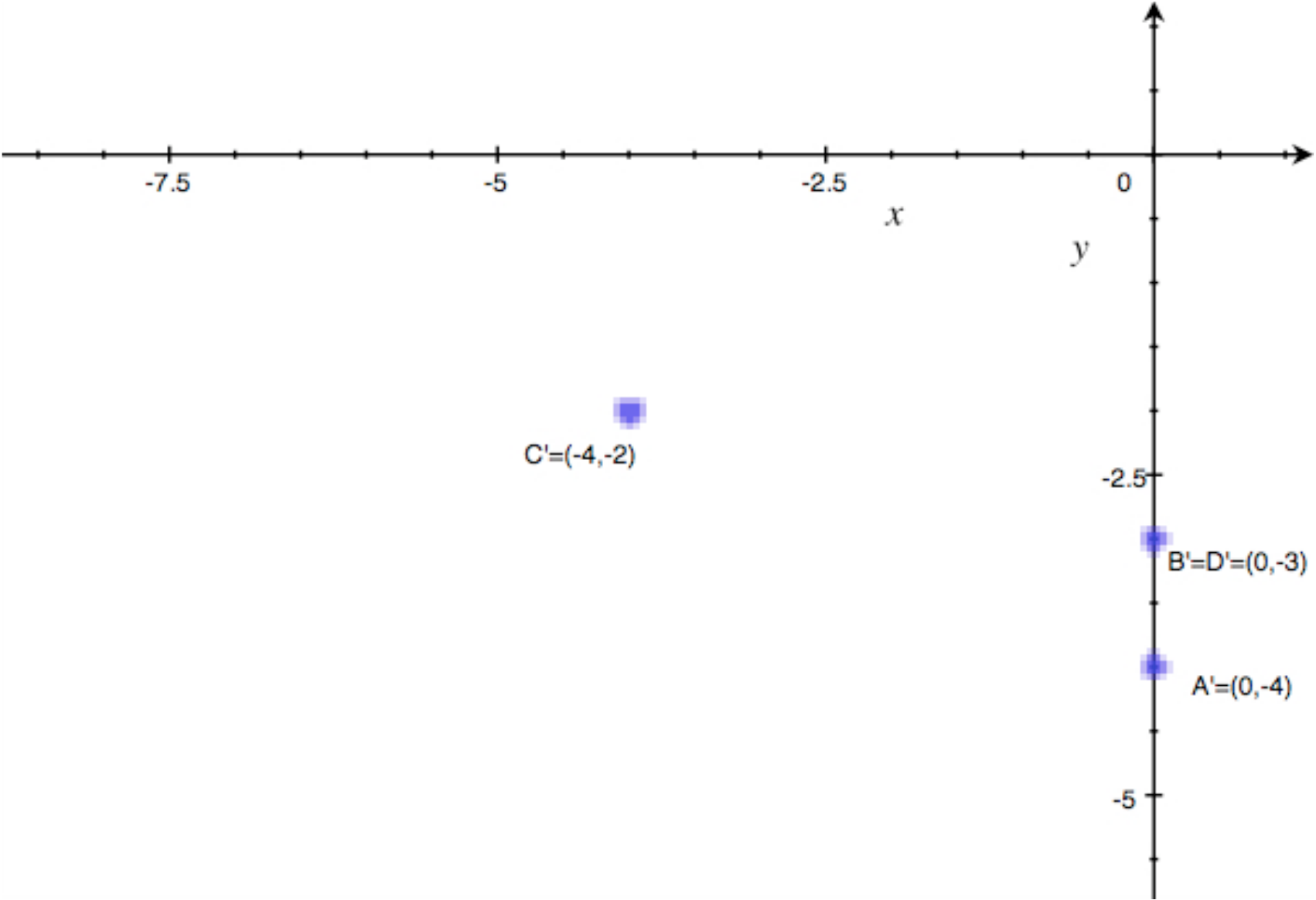} 
\caption{Initial finite payoff space, of the game $(M,<)$.}
\label{default}
\end{center}
\end{figure}

It is evident that the pair of strategies (E, L) is a dominant Nash equilibrium of the game, in other terms itÕs a strict Nash equilibrium (even if the Nash equilibria are two, also (N,L) is a Nash equilibrium). This Nash equilibrium leads to the gain payoff $(0,3)$. But, we have to do some considerations: 1) the second player could gain much more if the first does not enter; 2) the market offers a potential total gain 6, in correspondence to the bi-strategy (Enter, High prices). We have three questions: a) is it possible for the second player to gain more than 3? b) is it possible that the two firms collectively obtain the total amount offered by the market? c) if the case (b) happens, what is a possible fair division of the total amount among the firms? c) is it possible to enlarge the pie coopetitively? We shall answer to these questions during our study. To do so, first we consider the mixed extension of the game $(M,<)$.

\subsection{The mixed extension} 

\subsubsection{Scope of the mixed extension.} We shall examine the von Neumann extension of the finite game $(M,<)$ to find other possible realistic and applicable economic behaviors and solutions.

\subsubsection{The Extension.} We, firstly, have to imbed (canonically) the finite strategy spaces into the probabilistic canonical 1-simplex of the plane (since there are two strategies for any firm). The strategy Enter (and High prices) shall be transformed into the first canonical vector $e_1$ of the plane and the strategy Non Enter (and Low prices) shall be transformed into the second canonical vector of the plane $e_2$. So our new bistrategy space is the Cartesian square of the canonical 1-simplex of the plane (wich is the convex envelop of the canonical basis $e$ of the plane), $\mathrm{conv}(e)^2$. It is a 2-dimension bistrategy space, since the canonical 1-simplex is 1-dimensional. So we can imbed the simplex in the real line $\IR$ and the bi-strategy space into the Euclidean plane $\IR^2$. To do this, roughly speaking, we consider the injection associating with the pure strategies Enter and High prices the probability 1 and to the pure strategies Not Enter and Low prices the probability 0. In other (rigorous) terms, we associate to any mixed strategy $(x,1-x)$ the probability $x$ of the interval $[0,1]$ and to any mixed strategy $(y,1-y)$ the probability $y$ of the same interval. Moreover, the considered finite loss game $(M,<)$ is the translation by the loss vector $(0,-4)$ of the game $(M',<)$ represented in the following table:
\[
\begin{array}{ccc}

  	&	H	&  	L 	\\
  E	&	(-4,2)	&   	(0,1)	\\
  N	&	(0,1)	&   	(0,0)		   

\end{array}
\]

The mixed extension of the game $(M,<)$ is, thus, the translation, by the same vector $(0,-4)$, of the extension of the game $(M',<)$, so we can study this latter extension.

\subsection{The mixed extension and the coopetitive extension}

\textbf{Formal description of the mixed extension.} The mixed extension of the finite game $(M',<)$ is the infinite differentiable loss-game $G_0 = (f_0,<)$, with strategy sets $E = F = [0,1]$ and biloss (disutility) function $f_0$ defined, on the Cartesian square $\IU^2:=[0,1]^2$, by $f_0(x,y) = (-4xy,x+y)$, for every bistrategy $(x,y)$ in $\IU^2$. For what concerns the coopetition, we assume that the two firms decide to produce their products in a common industry, \emph{lowing the costs linearly} and so obtaining more gain. We obtain the new payoff function $f:\IU^3 \to \IR^2$ defined, for every bistrategy $(x,y)$ in $\IU^2$ and any cooperative strategy $z$ in the common set $C = \IU$, by $f(x,y,z) = (-4xy-z, x+y-z)$. The function $f$ is a coopetitive extension of $f_0$ and it is obtained by a \emph{coopetitive translation}, via the vector function $v:\IU \to \IR^2$, defined by $v(z) = -z(1,1)$, for every $z \in C$. The study of this two games $f_0,f$ could be conducted by analytical technics introduced and already applied by D. Carf\`{i} in \cite{ca1,ca2,ca3,ca4,ca5}. This study is briefly represented in the following figures, in which we show also some possible natural solutions.

\medskip

\textbf{In figure 2}, we see the payoff space of the game $f_0$. There are infinitely many Nash payoffs: the entire segment $[A',B']$. Note that in a cooperative perspective, all the minimal Pareto boundary under the Nash payoff $B'$ (the core of the game) are rationalizable. 
 
\begin{figure}[htbp]
\begin{center}
\includegraphics [width=7.5cm]{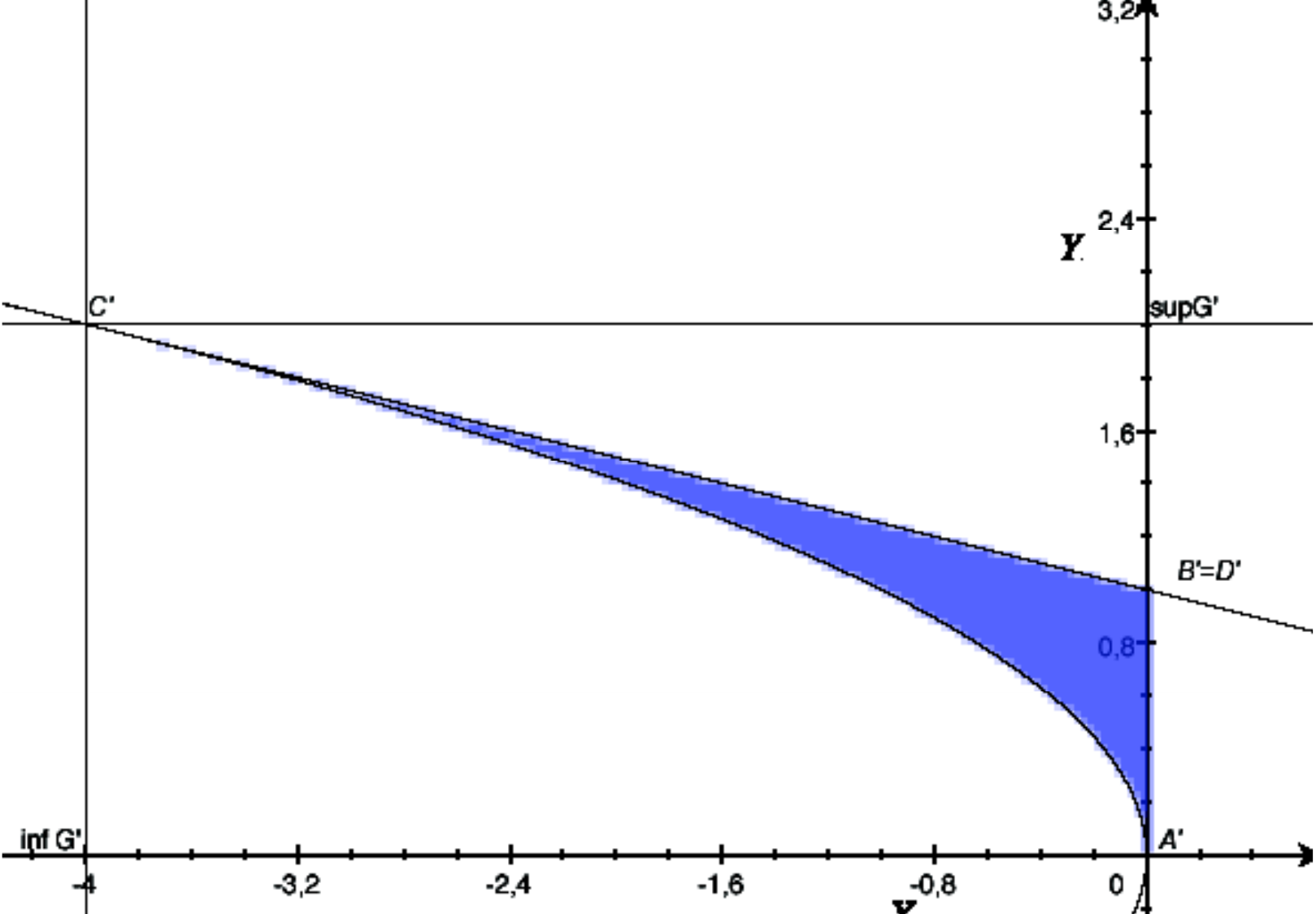} 
\caption{Payoff space of the mixed extension: $f_0 (\IU^2)$.}
\label{default}
\end{center}
\end{figure}

\medskip

\textbf{In figure 3}, we see two possible extended Kalai solutions $K'$ and $K''$ and the Nash bargaining solution $N'$.

\begin{figure}[htbp]
\begin{center}
\includegraphics [width=7.5cm]{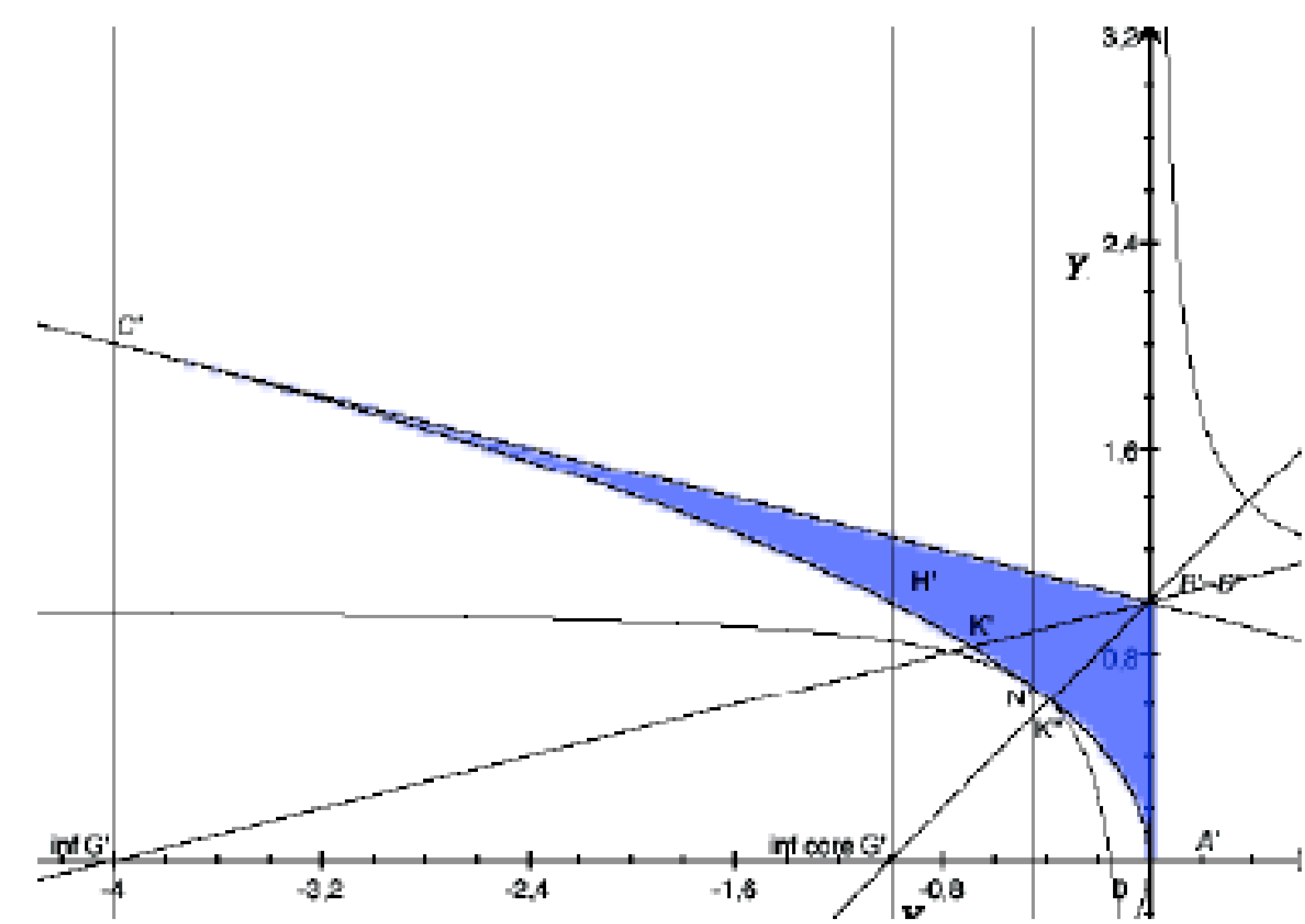} 
\caption{Two Kalai solutions $K',K''$ and the Nash bargaining solution $N$ on payoff space $f_0 (\IU^2)$.}
\label{default}
\end{center}
\end{figure}

\medskip

\textbf{In figure 4}, we see two possible extended transferable utility Kalai solutions $H$ and $K$, with respect to the conservative bi-value $B'$ and the infima of the the core and of the game.

\begin{figure}[htbp]
\begin{center}
\includegraphics [width=7.5cm]{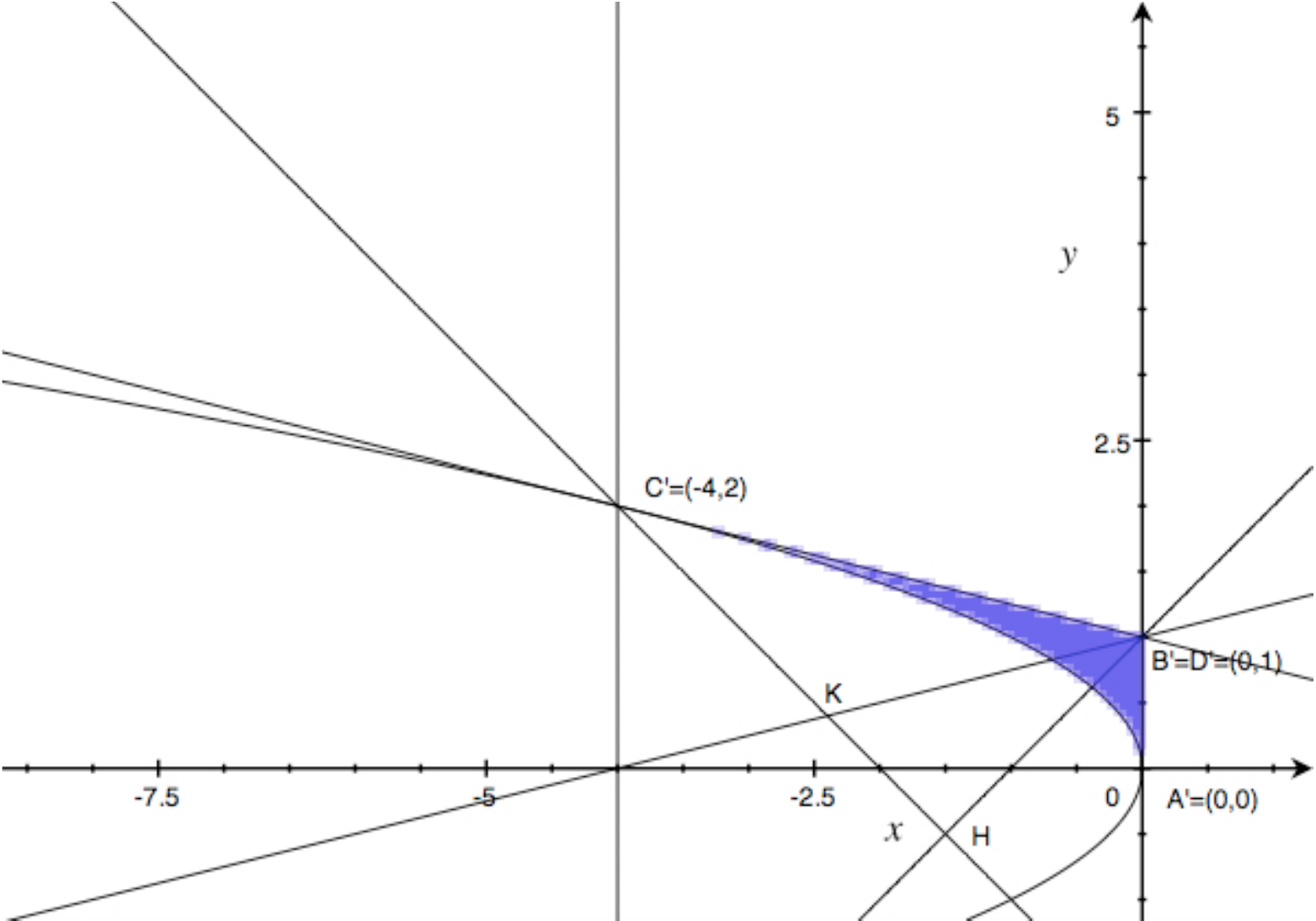} 
\caption{Two transferable utility Kalai-Smorodinsky solutions $H,K$ on payoff space $f_0 (\IU^2)$.}
\label{default}
\end{center}
\end{figure}

\medskip

\textbf{In figure 5}, we see the entire payoff space of the coopetitive game, that is the image $f(\IU^3)$.

\begin{figure}[htbp]
\begin{center}
\includegraphics [width=7.5cm]{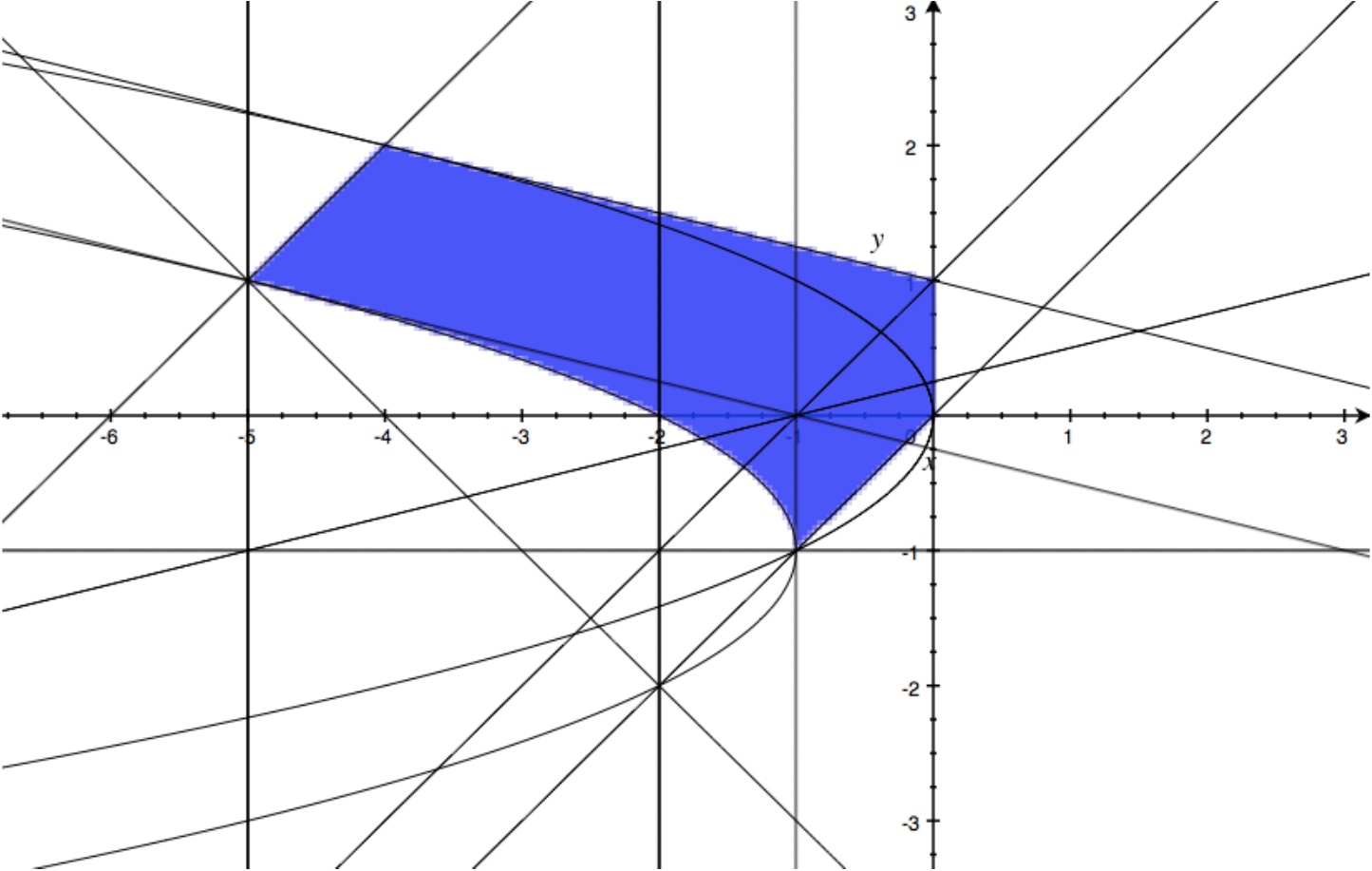} 
\caption{The coopetitive payoff space $f(\IU^3)$.}
\label{default}
\end{center}
\end{figure}

\medskip

\textbf{In figure 6}, we see two possible extended Kalai solutions $H'$ and $H''$, and two possible extended transferable utility Kalai solutions $K'$ and $K''$, having the conservative bi-value $B'$ (sup of the coopetitive Nash path) and the infimum of the partial  Nash coopetitive path $B' + [0,1]v(1)$ as threat points and the infimum of the entire coopetitive game as utopia point.

\begin{figure}[htbp]
\begin{center}
\includegraphics [width=7.5cm]{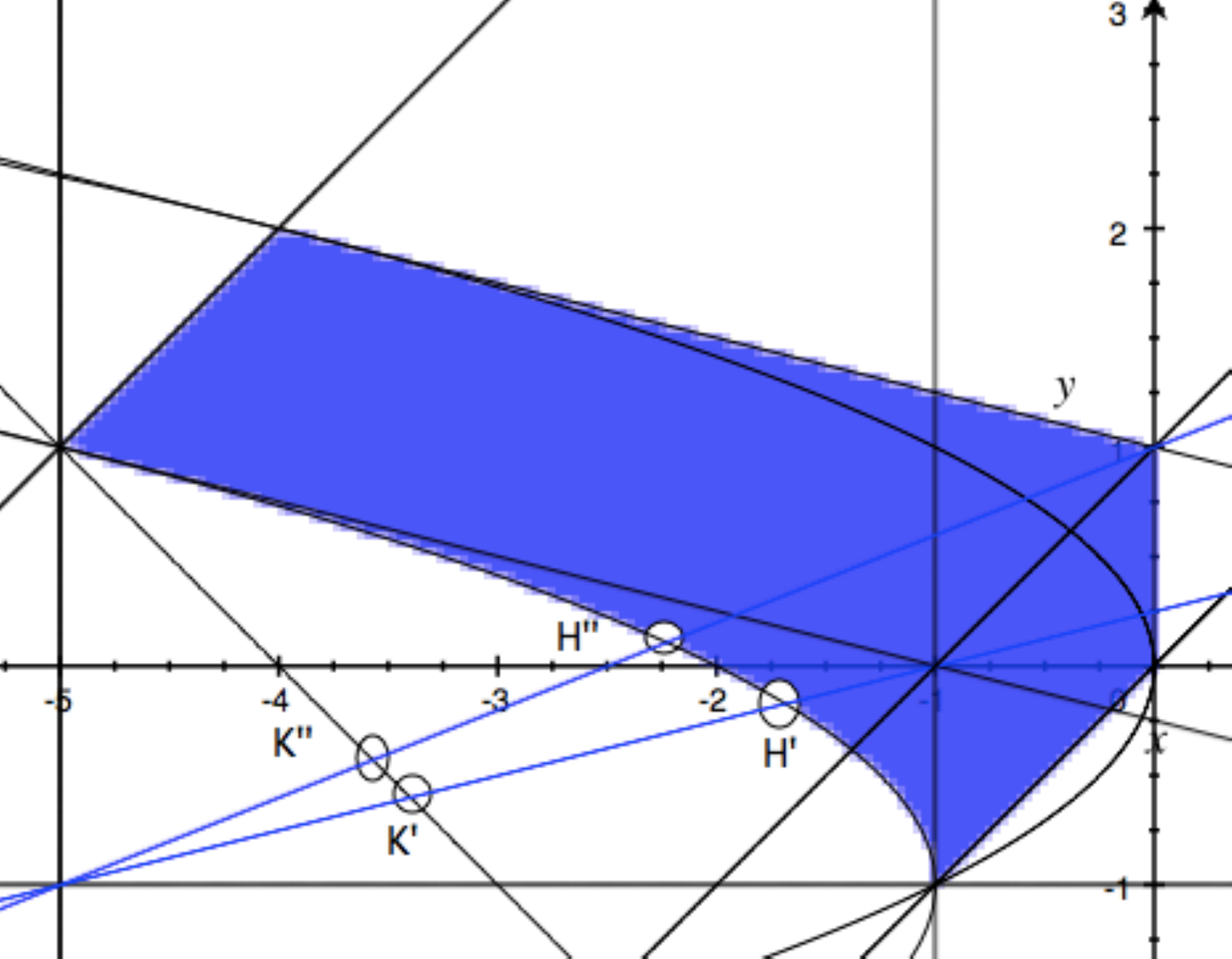} 
\caption{Kalai solutions $H',H''$ and transferable utility Kalai solutions $K',K''$ on coopetitive loss space.}
\label{default}
\end{center}
\end{figure}

\subsubsection{Conclusions.}
The sample of coopetitive game, provided in the present contribution, is essentially a \emph{normative model}. It has pointed out the strategies that could bring to win-win solutions, in a \emph{super-cooperative perspective and in a super-cooperative transferable utility perspective, after a coopetitive extension}. At this aim, we use an extended Kalai-Smorodinsky method, appropriate to determine fair partitions, for the \emph{win-win solutions}, on the Pareto boundary and transferable utility Pareto boundary of the coopetitive game. The solutions offered by our coopetitive model show, using Brandenburgher words, how to ``enlarge the pie and share it fairly''.

%\newpage

%\begin{small}

%\end{small}


\begin{thebibliography}{99}

\bibitem{ca1}  Carf\`{i} D., 2009. \textsl{Payoff space in $C^{1}$-games}, Applied Sciences (APPS), vol. 11, pp. 1-16.

\bibitem{ca2} Carf\`{i}, D., 2012. \textsl{A Model for Coopetitive Games}. MPRA paper 2012.

\bibitem{ca3} Carf\`{i} D., Perrone E., 2012. \textsl{Game complete analysis of symmetric Cournot duopoly}, MPRA paper.

\bibitem{ca4}  Carf\`{i} D., Schilir\`{o} D., 2011. \textsl{Crisis in the Euro area: coopetitive game solutions as new policy tools}, TPREF-Theoretical and Practical Research in Economic Fields, summer issue 2011, pg. 23-36.

\bibitem{ca5}  Carf\`{i} D., Ricciardello A. 2010. \textsl{An Algorithm for Payoff Space in C1-Games}, AAPP, LXXXVIII(1).


\end{thebibliography}
\end{document}